\documentclass[12pt,preprint]{aastex6}
\begin{document}

\title{Magnetar as Central Engine of Gamma-Ray Bursts: Quasi-Universal Jet, Event Rate and X-ray Luminosity Function of Dipole Radiations}
\author{Wen-Jin Xie, Le Zou, Hong-Bang Liu, Shan-Qin Wang, En-Wei Liang}

\affil{Guangxi Key Laboratory for Relativistic Astrophysics, School of Physical Science and Technology, Guangxi University, Nanning 530004, China; liuhb@gxu.edu.cn; lew@gxu.edu.cn\\}

\begin{abstract}
Early shallow-decaying X-ray afterglows of gamma-ray bursts (GRBs) may be attributed to the dipole radiations of newly-born magnetars. Assuming that the GRB jets powered by magnetars are quasi-universal, we find that the jet structure can be parameterized as a uniform jet with a luminosity of $\log L_{\rm j}/{\rm erg\ s^{-1}}=52.68^{+0.76}_{-0.33}$ (1$\sigma$) and an opening angle $\theta_{\rm j}=2.10_{-1.28}^{+1.90}$ (50\% confidence level), surrounding by a power-law decay component with an index of ${-4.00^{+0.27}_{-0.37}}$ (1$\sigma$). The inferred local GRB rate is $\rho=9.6$ Gpc$^{-3}$ yr$^{-1}$ by including both the typical GRBs and LL-GRBs as the same population. The typical viewing angle is $3.3^{o}$, and may be $20^{o}\sim30^{o}$ for LL-GRBs. The X-ray luminosity function of the dipole radiation wind can be empirically described by a broken power-law function with indices $\beta_1=0.78^{+0.16}_{-0.15}$ and $\beta_2>1.6$ broken at $\log L_{b, w}/{\rm erg\ s^{-1}}=48.51^{+0.53}_{-0.65}$. In case of that the wind outflow is collimated and co-axial with the GRB jet, we find that the wind structure is similar to the GRB jet, i.e., $\log L_{\rm c, w}/{\rm erg\ s^{-1}}=48.38^{+0.30}_{-0.48}$, $\theta_{\rm c, w}={2.65^{o}}_{-1.19^{o}}^{+0.1.73^{o}}$, and $k_{\rm w}=4.57^{+1.21}_{-0.75}$. The observed correlation between the prompt gamma-ray luminosity and X-ray luminosity of the wind may be resulted from the viewing angle effect in such a jet-wind system. Discussion on survey with the X-ray instruments on board the {\em Einstein\ Probe} mission in the soft X-ray band for the jet and wind emission is also presented.
\end{abstract}



\section{Introduction}

Gamma-Ray Bursts (GRBs) are the most violent explosions in the universe.
It is generally believed that GRBs are associated with death of massive stars or compact binary mergers (e.g., \citealp{2007ApJ...655L..25Z}; \citealp{2015PhR...561....1K}).
However, the central engine of GRBs is still inconclusive. There are two candidates that have been generally investigated, including a newly formed black hole (BH) with an accretion disk (\citealp{1992ApJ...395L..83N}; \citealp{1999ApJ...524..262M}; \citealp{2001ApJ...557..949N}) or a highly rotating magnetar (\citealp{1992Natur.357..472U}; \citealp{1994MNRAS.270..480T}; \citealp{1998A&A...333L..87D}; \citealp{2000ApJ...537..810W}; \citealp{2011MNRAS.413.2031M}; \citealp{2014ApJ...785...74L}).
In the hyper-accreting BH scenario, the relativistic GRB jets are powered through the Blandford-Znajek (BZ) mechanisms (\citealp{1977MNRAS.179..433B}) or by the polar annihilation of neutrino-antineutrino pairs (\citealp{1999ApJ...518..356P}).  For the magnetar scenario, the rotational energy of the magnetar could be extract to power the jet and product GRB via magnetic dipole radiation and relativistic wind spindown (\citealp{1992Natur.357..472U}; \citealp{1998A&A...333L..87D}), or could be extracted by magnetic bubble eruption due to differential rotation (\citealp{2006Sci...311.1127D}; \citealp{2000ApJ...542..243R}; \citealp{1998ApJ...505L.113K}).

 Burst Alert Telescope (BAT) and X-ray Telescope (XRT) on board $Swift$ satellite have detected thousands of GRBs and their X-ray afterglows. Half of well-sampled X-ray afterglow light curves observed by XRT are characterized with a shallow-decay segment or a plateau transiting to a normal-decay or a sharp drop segment (\citealp{2006ApJ...642..389N}; \citealp{2006Natur.444.1010Z}; \citealp{2006ApJ...647.1213O}; \citealp{2007ApJ...670..565L}; \citealp{2007ApJ...665..599T}; \citealp{2014ApJ...785...74L}; \citealp{2016MNRAS.462.2990D}). A newly-born magnetar is thought to be the central engine of a long-lasting X-ray plateau since the injected kinetic luminosity of the magnetic dipole (MD) radiation is steady during the characteristic spin-down timescale (e.g.,  \citealp{1998A&A...333L..87D}; \citealp{2001ApJ...552L..35Z}). This scenario has been supported by $Swift$  observations for both long and short GRBs (\citealp{2006MNRAS.372L..19F}; \citealp{2007ApJ...670..565L}; \citealp{2007ApJ...665..599T}; \citealp{2010MNRAS.402..705L}; \citealp{2013MNRAS.430.1061R}; \citealp{2011A&A...526A.121D}; \citealp{2013MNRAS.430.1061R}; \citealp{2013MNRAS.431.1745G}; \citealp{2014MNRAS.438..240G}; \citealp{2014ApJ...785...74L}; \citealp{2015ApJ...805...89L}; \citealp{2016PhRvD..93d4065G}; \citealp{2017MNRAS.470.4925G}; \citealp{2018MNRAS.478.4323G}). Recently, \cite{2019Natur.568..198X} reported a redshift-known transient with X-ray plateau, which is thought to be powered by a newly-born magnetar as the aftermath of a binary neutron-star merger.
It has been widely studied for the GRB jet luminosity function (\citealp{2001ApJ...552...36S}; \citealp{2005ApJ...619..412G}; \citealp{2007ApJ...662.1111L}; \citealp{2010MNRAS.406.1944W}; \citealp{2014ApJ...783...24L}).
However, the X-ray luminosity function of newly-born magnetar dipole radiation has been rarely explored before, \cite{2017ApJ...835....7S} perform peak luminosity function and the event rate of the X-ray transients associated with double neutron star mergers. Determining the intrinsic distribution of dipole radiation luminosity and describing it by equation would provide key clues to understand the properties of newly-born magnetars and provide the detection guidance for the future X-ray telescope mission.

Laser Interferometer Gravitational-wave Observatory (LIGO) detected GW 170817 located in a nearby galaxy whose  distance is 40 Mpc in 2017 (\citealp{2017ApJ...848L..12A}). Two seconds later, a weak gamma ray burst GRB 170817A associated with GW170817 was observed independently by Gamma-ray Burst Monitor (GBM) on board $Fermi$.
Models of central engines powering such relativistic jets are still in debate.
Recent researches (e.g., \citealp{2017ApJ...848L..20M}; \citealp{2019MNRAS.483..840B}) suggested that the low luminosity GRB 170817A may be produced by an off-axis structured jet.
Those researches indicated that at least a fraction of LL-GRBs are off-axis ones.
The multi-messenger and multi-band observations (\citealp{2017PhRvL.119p1101A}) provide an unprecedent opportunity to explore the structure of the GRB jet.
The prompt emission is believed to be strongly beamed along the ultra-relativistic jets with half-opening angles $\theta_j$ (\citealp{1999ApJ...525..737R}; \citealp{1999ApJ...519L..17S}; \citealp{2001ApJ...562L..55F}).
Three jet structure are generally discussed in the literatures, including a uniform jet, a power-law (\citealp{1998ApJ...499..301M}; \citealp{2001ApJ...552...72D}; \citealp{2002ApJ...571..876Z}) and a gaussian structure jet (\citealp{2002ApJ...571..876Z}; \citealp{2003ApJ...591.1075K}).
\cite{2004ApJ...601L.119Z} showed that the GRB-XRF prompt emission/afterglow data can be explained in the frame of quasi-universal structure jet and derived a typical opening angle 5.7 deg  with a standard energy about $10^{51}$ erg.
 The local rate of high-luminosity GRB (HL-GRB) is found to be $\sim1$ Gpc$^{-3}$ yr$^{-1}$ (\citealp{2001ApJ...552...36S}; \citealp{2007ApJ...662.1111L}; \citealp{2010MNRAS.406.1944W}; \citealp{2014ApJ...783...24L}). However, the Low Luminosity GRB (LL-GRB) local rate is nearly $100\sim1000$ Gpc$^{-3}$ yr$^{-1}$ (\citealp{2006ApJ...645L.113C}; \citealp{2006Natur.442.1011P}; \citealp{2007ApJ...662.1111L}). Based on the distinction of local rate, luminosity and beaming factor between HL-GRB and LL-GRB. Some groups (e.g., \citealp{2007ApJ...662.1111L}; \citealp{2012ApJ...747...88N}; \citealp{2011ApJ...739L..55B}) suggested the LL-GRBs are likely a unique GRB population.
 Other groups (e.g., \citealp{1999ApJ...522L.101N}; \citealp{2003Natur.426..154B}; \citealp{{2003ApJ...594L..79Y}}) argued that LL-GRBs and HL-GRBs are common origin but observed from different angles.

In this paper, we explore whether the sample of those GRBs whose early X-ray afterglow lightcurves are characterized by a shallow-decaying segment can be reproduce by a quasi-universal structure jet from different viewing angles. We constrain the parameters of the jet structure through Monte Carlo simulations by comparing the observed luminosity and redshift distributions to the simulation results, and revisit the GRB rate in this scenario. With the same technique, we also constrain the luminosity function (LF) of the dipole radiation winds powered by the magnetars and explore possible outcome of X-ray survey mission in the near future. This paper is organized as follows. In Section 2, we present the data and our GRB sample selection. The constrain on the jet structure and event rate are shown in Section 3, and constraints on the LF of the dipole radiation wind are reported in Section 4.
Discussion and conclusion are drawn in Section 5 and Section 6, respectively.
 A flat Universe with $H_0=71.0$ km s$^{-1}$Mpc$^{-1}$, $\Omega_M$=0.3 and $\Omega_\Lambda$=0.7 is adopted throughout this paper.

\section{Sample Selection}
\label{Sec:sample}
We adopt a GRB sample presented in \cite{2019ApJ...877..153Z} for our analysis. These GRBs are selected from current Swift GRBs with redshift measurement. The X-ray lightcurves of these GRBs have an early X-ray shallow-decay segment or plateau with a flux decay index\footnote{The decay index of emission from forward shocks should be steeper that 0.75 in the framework of the standard external shock model without considering late energy injection (e.g., \citealp{2007ApJ...670..565L}; \citealp{2008ApJ...675..528L}).} as $\alpha < 0.75$, where $ F \propto t^{-\alpha}\nu^{-\beta}$ is adopted. No bright X-ray flare is  observed in the shallow decay segment (plateau).

It was proposed that the shallow-decay segment (plateau) is attributed to emission of a long-lasting, steady wind powered by the dipole radiations of a newly-born magnetar (e.g., \citealp{1998A&A...333L..87D}; \citealp{2001ApJ...552L..35Z}; \citealp{2007ApJ...670..565L}). Thus, the selected GRBs are regarded as a sample of GRBs powered by newly-born magnetars. It was suggested that the central engines of GRBs 980425, 060218, 170817A like low-luminosity GRBs (LL-GRBs) are also a magnetar (\citealp{2000ApJ...537..810W}; \citealp{2007ApJ...659.1420T}; \citealp{2017ApJ...851L..18W}). We therefore include LL-GRBs 980425, 060218, and 170817A in our sample.

The unabsorbed BAT and XRT light-curve data of these GRBs are taken from the {$Swift$} website (\citealp{2010A&A...519A.102E})\footnote{$http://www.swift.ac.uk/burst$ $analyser/$}.  To make joint X-ray light curve with data observed by BAT and XRT, the BAT data are extrapolated to the XRT band (0.3-10 KeV) by assuming a single power-law spectrum (\citealp{2006ApJ...647.1213O}; \citealp{2007A&A...469..379E}; \citealp{2009MNRAS.397.1177E}). We also collect the 1-s peak photon flux and photon indices of both the prompt gamma-ray and X-ray plateau phases for these GRBs from the website. It is well known that GRB spectra are usually well fitted with the so-called Band function, which is smooth broken power-law with a peak energy $E_{\rm p}$ (\citealp{1993ApJ...413..281B}). However, the spectra of prompt gamma-rays observed with BAT are usually fitted with a single power-law (SPL) model, $F\propto E^{-\Gamma}$), being due to the narrowness of the BAT band (e.g., \citealp{2008ApJS..175..179S}). \cite{2012MNRAS.424.2821V} reported an empirical relation,
\begin{equation}
\log(E_{\rm peak})=(4.34\pm 0.475)-(1.32\pm 0.129)\Gamma^{\rm BAT}.
\label{eq:Ep-Gamma}
\end{equation}
We adopt this empirical relation to estimate the $E_p$ values of these GRBs and calculate the bolometric luminosity in the $1-10^{4}$ keV band with the Band function adopting the typical photon spectral indices $ \alpha=-1 $ and $ \beta=-2.3 $ for all selected GRBs.
We make empirical fit to the shallow decay segment in the early XRT lightcurves to measure the end time ($t_b$) and the corresponding flux ($F_w$) of this segment. We take the characteristic X-ray luminosity of dipole radiation wind as $L_{\rm w}=4 \pi D_{L}^2 F_{w}$. The data of the GRBs in our sample are reported in Table 1.

\section{GRB Jet Structure and local event rate}
\subsection{Methodology}
We parameterize the jet structure as
(\citealp{2015MNRAS.447.1911P},\citealp{2018MNRAS.473L.121K})
\begin{equation}\label{jet_structure}
L_{\rm j}(\theta|k, \theta_{\rm c, j}, L_{\rm c,j})= \left\{
 \begin{array}{ll}
  L_{\rm c,j} & \theta \leq \theta_{\rm c, j},
   \nonumber \\
   L_{\rm c,j}(\frac{\theta}{\theta_{\rm c, j}})^{-k_{\rm j}} & \theta > \theta_{\rm c, j},
 \end{array}
\right.
\end{equation}
where $L_{\rm c,j}$ is isotropic luminosity of the jet core within $\theta_{\rm c, j}$, and $k_{\rm j}$ is the power-law decaying index out of the core. The number density of GRBs in unit time at unit redshift is given by
\begin{equation}\label{dn}
\frac{dN}{dtdz}=\frac{R_{\rm GRB}(z)}{1+z}\frac{dV(z)}{dz},
\end{equation}
where $R_{\rm GRB}(z)$ is the GRB event rate in the unit of ${\rm Gpc^{-3} yr^{-1}}$ at a comoving volume element $\frac{dV(z)}{dz}$, which is calculated with
\begin{equation}\label{dvdz}
\frac{dV}{dz}=\frac{c}{H_0}\frac{4\pi D^2_L}{(1+z)^2
[\Omega_M(1+z)^3+\Omega_\Lambda]^{1/2}}.
\end{equation}
The factor $(1+z)^{-1}$ accounts for the cosmological time dilation correction.
The GRB rate is assumed to follow the star formation rate.
We also consider the enhanced evolution factor of GRBs (\citealp {2008ApJ...683L...5Y}; \citealp{2009ApJ...705L.104K}; \citealp{2009ApJ...705L.104K}; \citealp{2010MNRAS.406..558Q} ), i.e., the GRB rate $R_{GRB}(z)$ is parameterized as (\citealp {2008ApJ...683L...5Y} )
\begin{equation} \label{GRB_rate}
R _{\rm LGRB}(z) \propto R_{\rm SFR} (z) \mathcal{E}(z),
\end{equation}
where $R_{\rm SFR}(z)$ is the star formation rate at redshift $z$, $\mathcal{E}(z)=\mathcal{E}_{0}(1+z)^{\alpha}$ with $\alpha \simeq 1.5$ stands for an enhanced evolution of GRBs.
There are many previous studies for the star formation rate $R_{\rm SFR}(z)$, (see, e.g., \citealp{1999Ap&SS.266..291R}; \citealp{2006ApJ...651..142H}; \citealp {2008ApJ...683L...5Y}; \citealp{2010MNRAS.406..558Q}). The equation form and parameters presented by \cite {2008ApJ...683L...5Y} are adopted in our analysis, which is

\begin{equation}
R_{\rm SFR}(z)={R}_{0}\left[(1+z)^{a \eta}+\left(\frac{1+z}{B}\right)^{b \eta}+\left(\frac{1+z}{C}\right)^{c \eta}\right]^{1 / \eta},
\end{equation}
where $R_0$ is the local SFR rate, $a=3.4$, $b=-0.3$, $c=-3.5$, $B=5000$, $C=9$, and $\eta=-10$.

\subsection{Simulation analysis}
We constrain the jet parameters of $k_{\rm j}$, $\theta_{\rm c, j}$, $L_{\rm c,j} $ via Monte Carlo simulation analysis by comparing our simulations to the data observed with the {\em Swift} mission. We assume uniform distributions of these parameters in the range $k_{\rm j}\in \{2,10\}$, $\theta_{\rm c, j}\in\{0.5^{o},20^{o}\}$, and $\log L_{\rm c,j}/{\rm erg}\in \{51,\ 54\}$. We generate a jet structure parameter set of $\{L_{\rm c,j}, \theta_{\rm c, j}, k_{\rm j}\}$ by randomly picking up their values from their prior uniform distributions. The redshift ($z^{i}$) of a given mock GRB $i$ is generated via a bootstrap method based on Eqs. (\ref{GRB_rate}). Since the probability of observing a GRB in $\theta_v$ is given by
\begin{equation}\label{P_theta}
p(\theta_v)\propto \sin \theta_v,
\end{equation}
we generate its viewing angle $\theta^{i}$ based on Eq. (\ref{P_theta}). To avoid over producing GRBs with luminosity of $L_{\rm c,j}$ for $\theta_v<\theta_{\rm c, j}$, we further assume that $\log L_{\rm c,j}$ is quasi universal, being a Gaussian distribution with $\sigma_{L_{\rm c,j}}=0.5$, i.e.,
\begin{equation}\label{Lc}
\Phi\left(L^{i}_{\rm c, j}\right)\propto \exp \left[-\frac{\left(\log L^{i}_{\rm c, j}-\log L_{\rm c,j}\right)^{2}}{2 \sigma_{\log L_{\rm c,j}}^{2}}\right]
\end{equation}
for a given GRB $i$. We generate an $L^{i}_{\rm c, j}$ value via a bootstrap method based on Eq. (\ref{Lc}). Its observed luminosity viewing at $\theta^{i}_v$, $L^{i}(\theta^{i}_v)$, is derived from Eq. (\ref{jet_structure}) with the parameter set  $\{L^{i}_{\rm c, j}, \theta_{\rm c, j}, k_{\rm j}\}$. Hence, its observed flux is calculated with
\begin{equation}
F^{i}_{\rm obs}=\frac{L^{i}(\theta^{i}_v)} {4\pi D^{2}_L(z^{i}) k_{\rm corr}},
\end{equation}
where $D_L(z^{i})$ is the luminosity distance of the GRB $i$ and $k_{\rm corr}$ is the k-correction factor for correcting the bolometric flux to the flux at a given instrument energy band. We adopt $k_{\rm corr}=3$ in our simulation for a typical GRB with the Band function spectral parameters of $E_p=200$ keV, $\Gamma_1=-1$, $\Gamma_2=-2.3$ at $z=2$. In case of $F^{i}_{\rm obs}$ is higher than the BAT instrument threshold, i.e., $F^{i}_{\rm obs}>F_{th}$, this event is picked up as a triggered GRB. The lowest flux truncation of BAT is $1\times 10^{-8}$ erg cm$^{-2}$ s$^{-1}$ for GRBs with an incident angle of zero, but it is lowered down to $1\times 10^{-7}$ erg cm $^{-2}$ s $^{-1}$ for GRBs with an incident angle of $55^o$ (\citealp{2014ApJ...783...24L}). Since the incidence angle of a GRB is random, the trigger criterion would be uniform in the range $\{1\times 10^{-8}$, $1\times 10^{-7}$\} erg cm$^{-2}$ s$^{-1}$. Therefore, we pick a random value in this range as the trigger criterion for a given mock GRB.
We simulate a mock {\em Swift} GRB sample of 1500 GRBs (comparable to the current {\em Swift} GRB sample) for each parameter set $\{L_{\rm c,j}, \theta_{\rm c, j}, k_{\rm j}\}$, and evaluate the consistency between the observed and mock GRB sample of GRB distributions in the $L_{\rm obs}-z$ plane, using the probability of the Kolmogorov-Smirnov test (K-S test), $p_{\rm KS}\equiv p^{L}_{\rm KS}\times p^{z}_{\rm KS}$, where $p^{L}_{\rm KS}$ and $p^{z}_{\rm KS}$ are the probabilities of the K-S tests for the luminosity and redshift distributions.

\begin{figure}
\plotone{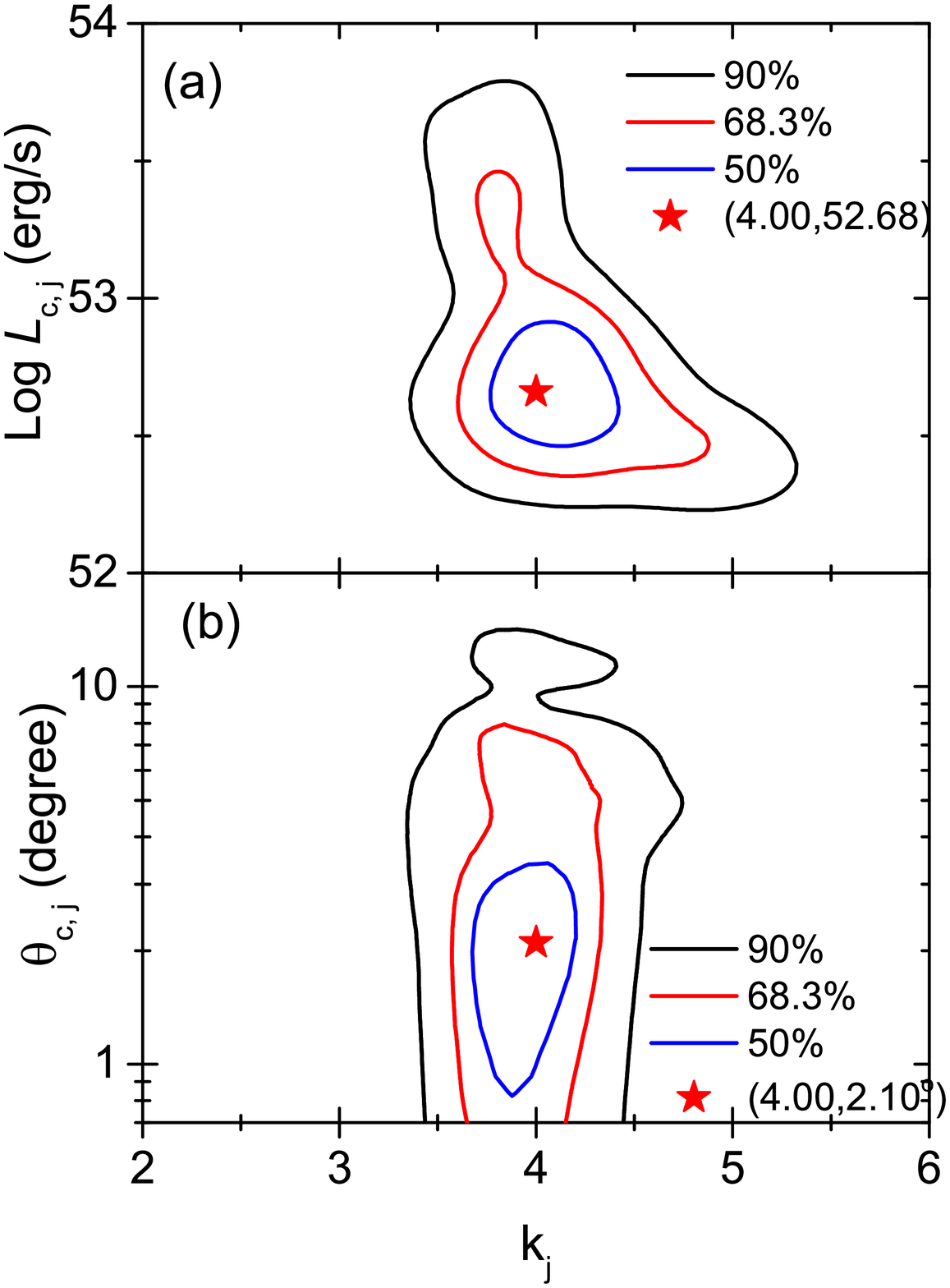}
\caption{Contours of the Kolmogorov-Smirnov test value $p_{\rm KS}$ ($p_{\rm KS}\equiv p^{L}_{\rm KS}\times p^{z}_{\rm KS}$) distribution in the $k_{\rm j}-\log L_{\rm c,j}$ and $k_{\rm j}-\theta_{\rm c,j}$ planes, where $L_{\rm c,j}$ is the isotropic luminosity within $\theta_{\rm c, j}$ and $k_{\rm j}$ is the power-law decaying index out of $\theta_{\rm c, j}$. The confidence level contours of 50\%, 68.3\% and 90\% are plotted by normalized the $P_{\rm KS}$ to its maximum value (the red stars).
\label{fig:jet-structure}}
\end{figure}

We simulate $2\times 10^4$ sets of $\{L_{\rm c,j}, \theta_{\rm c, j}, k_{\rm j}\}$. Figure \ref{fig:jet-structure} shows the $P_{\rm KS}$ contours in the $L_{\rm c,j}-k_{\rm j}$ and $\theta_{\rm c, j}-k_{\rm j}$ planes. The derived best parameter set is $\{\log L_{\rm c,j}, \theta_{\rm c, j}, k_{\rm j}\}=\{52.68, 2.1^{o}, 4.00\}$ with a $P^{\rm max}_{\rm KS}=0.19$. We normalize the $P_{\rm KS}$ value of each parameter set to $P^{\rm max}_{\rm KS}$, then derive the confidence level contours of 50\%, 68.3\% and 90\%. Constraints on $L_{\rm c, j}$ and $k_{\rm j}$ in a confidence level of $68.3\%$ ($1\sigma$) are $\log L_{\rm c, j}/{\rm erg/s}=52.68^{+0.76}_{-0.33}$ and $k_{\rm j}=4.00^{+0.27}_{-0.37}$. An upper limit of $\theta_{\rm c, j}<8^{o}$ is found in $1\sigma$ confidence level, and it is $\theta_{\rm c, j}=2.10_{-1.28}^{+1.90}$ in a confidence level of $50\%$.

\begin{figure}
\plotone{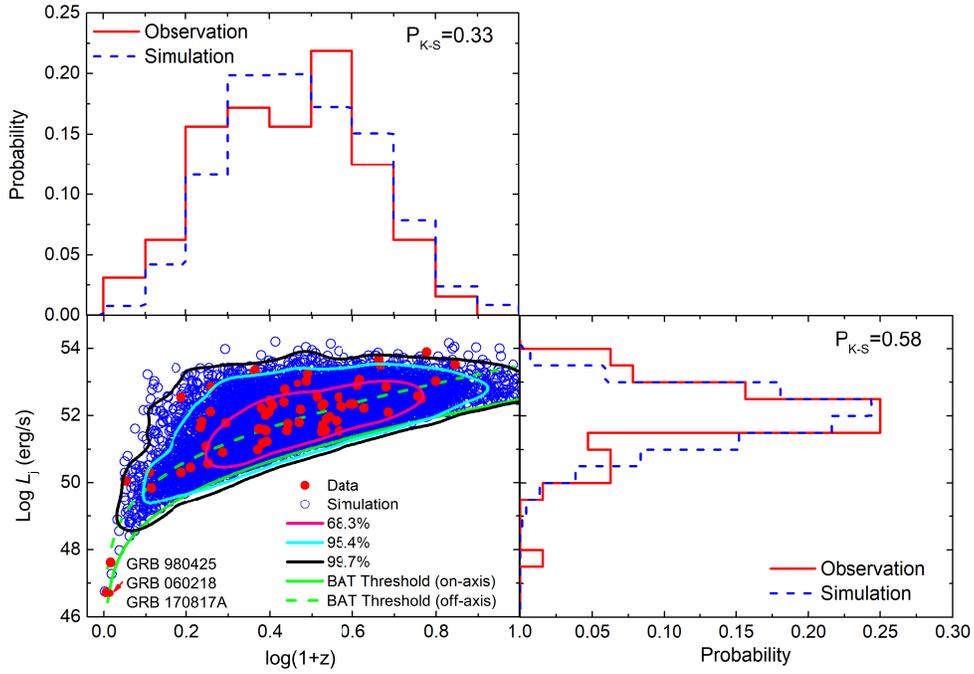}
\caption{
Comparison of the observed sample (solid dots) and mock sample (opened dots) derived from the best parameter set $\{\log L_{\rm c,j}/{\rm erg\ s^{-1}}=52.68\pm 0.50, \theta_{\rm c, j}=2.10^\circ, k_{\rm j}=4.00\}$ in the 1 and 2-dimensional $\log (1+z)-\log L_{\rm j}$ plane. The contours of $68.3\%$,  $95.4\%$, and $99.7\%$ of the mock GRB distirbution are also shown. The $Swift$/BAT sensitivity curves for events with extremely large incidence angle ($\sim 55^{o}$). $F^{\rm BAT}_{\rm th}=1 \times 10^{-7}$erg s$^{-1}$ cm$^{-2}$ and for events with extremely small incidence angle $F^{\rm BAT}_{\rm th}=1\times 10^{-8}$erg s$^{-1}$ cm$^{-2}$ are indicated with solid and dashed lines, respectively.
\label{fig:jet-1D-2D}}
\end{figure}

\begin{figure}
\plotone{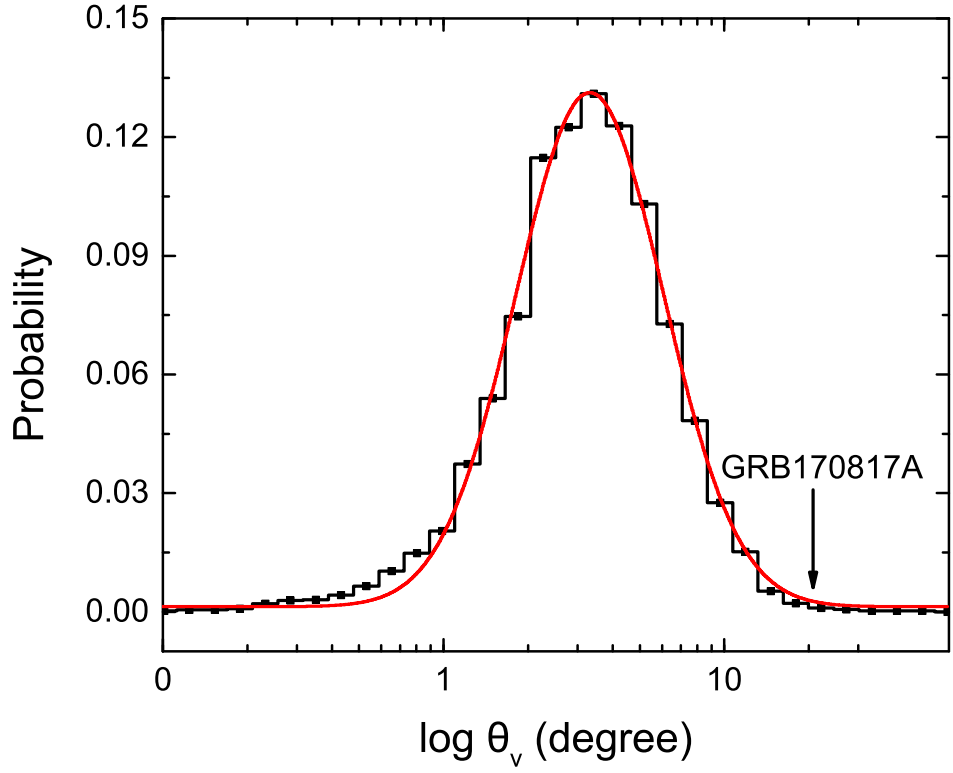}
\caption{Distribution of viewing angles derived from simulation based on the best parameters $\{\log L_{\rm c,j}/{\rm erg\ s^{-1}}=52.68\pm 0.05, \theta_{\rm c, j}=2.10^\circ, k_{\rm j}=4.00\}$.
\label{fig:viewing-angle}}
\end{figure}

Figure \ref{fig:jet-1D-2D} shows comparison of the observed sample to the mock sample derived from our simulations based on the best parameter set $\{\log L_{\rm c,j}, \theta_{\rm c, j}, k_{\rm j}\}=\{52.68, 2.1^{o}, 4.00\}$. One can observe that the data can be well reproduced with $P^{L}_{\rm KS}=0.58$, $P^{z}_{\rm KS}=0.33$, and $P_{\rm KS}=0.19$. The derived distribution of viewing angles is shown in Figure \ref{fig:viewing-angle}. It can be well fit with a log-normal function of $\log \theta_v=0.52\pm 0.13$. A few nearby LL-GRBs with a luminosity of several $10^{46}$ erg $s^{-1}$, such as GRBs 980425 ($z=0.0086$), 060218 ($z=0.0331$), and 170817A ($z=0.0098$) are indeed can be reproduced. They are $\log L/(\rm erg s^{-1})=46.57$ at $z=0.00618$ and $\log L/(\rm erg s^{-1})=46.66$ at $z=0.01533$. Their viewing angles are 32$^{o}$ and 28$^{o}$, being similar to that of GRB 170817A. The probability of such events is only about $10^{-4}$.

We evaluate the GRB event rate based on the derived best parameter set of the jet structure. The number of the detected GRBs with an instrument having a flux threshold
$F_{\rm th}$ and an average solid angle $\Omega$ for the aperture
flux in a period of $T$ should be calculated with
\begin{equation}\label{N}
N=\frac{\Omega T}{4\pi}\int_{L_{c, \min}}^{L_{c, \max}}
\Phi(L_{\rm c,j})dL_{\rm c,j} \int_{0}^{\theta_{v, \max}} p(\theta_v)d\theta_v  \int_0^{z_{\max}} \frac{R_{\rm
GRB}(z)}{1+z}\frac{dV(z)}{dz}dz,
\end{equation}
where $L_{c, \max}$ and $L_{c, \min}$ are taken as $10^{55}$ and $10^{50}$ erg s$^{-1}$,
respectively, and $\theta_{c, \max}$ and $z_{\max}$ depend on the jet structure (Eq. 5) and the
instrumental flux threshold $F_{\rm th}$. The solid angle of BAT is 1.33, and the average number of {\rm Swift}/BAT is $\sim$95 per year based on observations. Adopting the best parameter set $\{\log L_{\rm c,j}, \theta_{\rm c, j}, k_{\rm j}\}=\{52.68, 2.1^{o}, 4.00\}$ and $F^{\rm BAT}_{\rm th}=3\times 10^{-8}$ erg cm$^{-2}$ s$^{-1}$, we obtain the local GRB rate is $\rho_0=9.6$ Gpc$^{-3}$yr$^{-1}$ by riskily assuming that all GRBs have such a jet structure.

\section{X-Ray Luminosity Function and Wind Structure of Newly-born Magnetar Dipole Radiations}
{\em Swift}/XRT may present the first sample of newly-born magnetar dipole radiations in the X-ray band with its prompt slewing capacity (several tens of seconds post the BAT trigger), high sensitivity (with a threshold of $10^{-13}$ erg cm$^{-2}$ s$^{-1}$ in 0.3-10 keV), and long operation years (about 14 years since {\em Swift} launch in 2004). We constrain on X-ray luminosity function and wind structure of newly-born magnetar dipole radiations with this sample.

\begin{figure}
\plotone{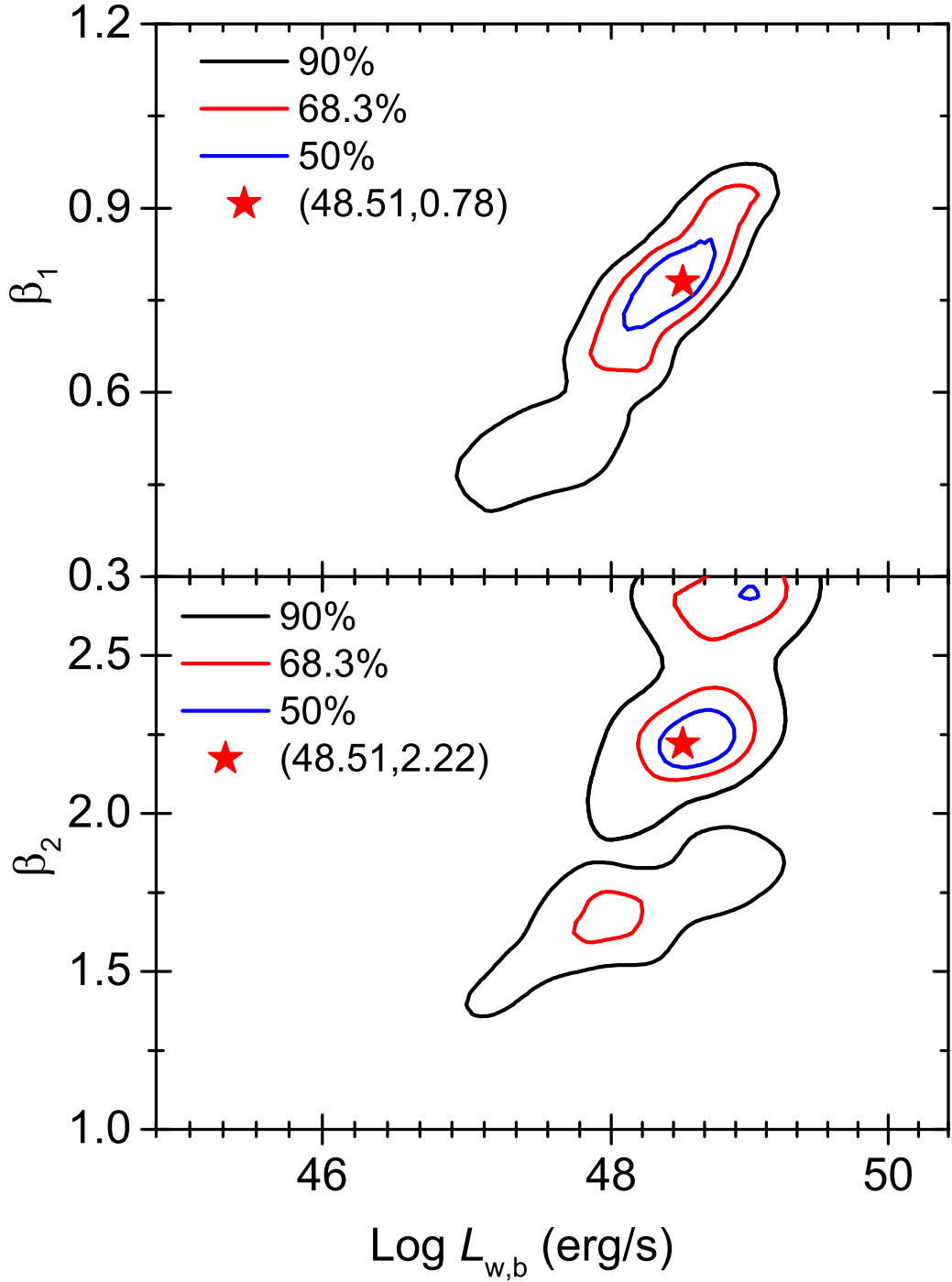}
\caption{Contours of the Kolmogorov-Smirnov test value $p_{\rm KS}$ ($p_{\rm KS}\equiv p^{L}_{\rm KS}\times p^{z}_{\rm KS}$) distribution in the $\beta_1-\log L_{\rm w, b}$ and $\beta_2-\log L_{\rm w, b}$ planes, where $\beta_1$ and $\beta_2$ are the indices before the break luminosity $L_{\rm w, b}$ of the empirical luminosity function for the wind emission. The confidence level contours of 50\%, 68.3\% and 90\% are plotted by normalized the $P_{\rm KS}$ values to their maximum value (the red stars).}
\label{fig:wind-LF-parameter}
\end{figure}

\begin{figure}
\plotone{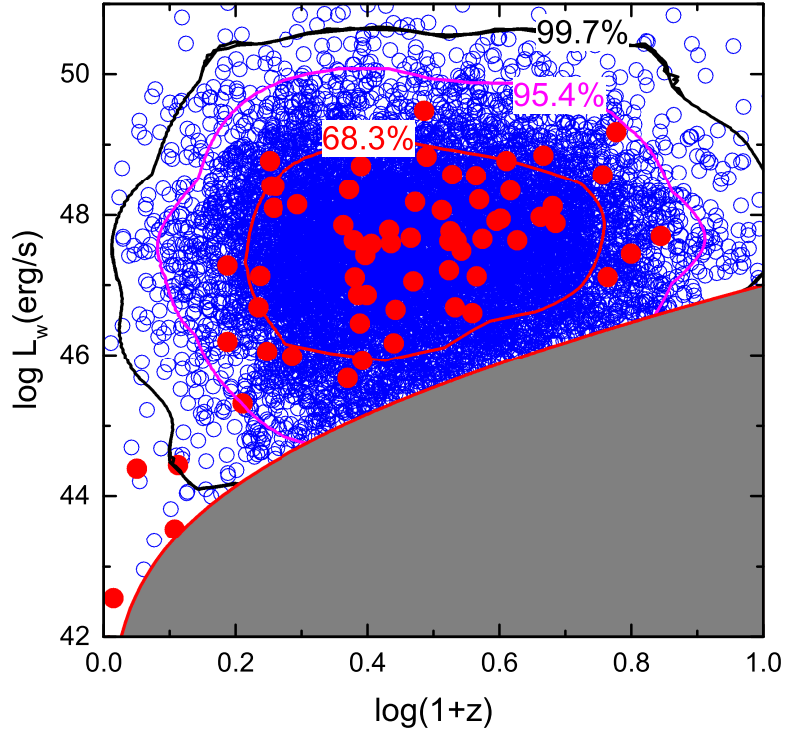}
\caption{Comparison between the observed sample (solid dots) and the probability distribution contours derived from our simulations based on the best parameter set $\{\log L_{\rm w, b}/{\rm erg\ s^{-1}}=48.51, \beta_{1}=0.78, \beta_2=2.22\}$. The solid red line stands for the roughly estimated sensitivity of $Swift$ XRT ($1\times10^{-13}$ erg cm$^{-2}$ s$^{-1}$).}
\label{fig:wind-LF-Dis}
\end{figure}

We first investigate the empirical luminosity function by characterize it with a broken power-law function,
\begin{equation}\label{LF}
\Phi(L_{\rm w})=\Phi_0\left[\left(\frac{L_{\rm w}}{L_{b, w}}\right)^{\beta_1}
+\left(\frac{L_{\rm w}}{L_{b, w}}\right)^{\beta_2}\right]^{-1}.
\end{equation}
The indices $\beta_1$ and $\beta_2$ as well as the broken luminosity $L_{b, w}$ are constrained by measuring the observed $L^{\rm obs}_{w}$ and $z$ distributions with the our simulation results using the same method as above. The flux threshold of XRT in blind search mode is taken as $F_{\rm XRT, th}=1\times10^{-13}$ erg cm$^{-2}$ s$^{-1}$. Our procedure is the same as that for constraining the jet structure parameters as described above. The $L_{\rm w}$ values are bootstrapped based on Eq. \ref{LF} in the range of $\log L_{b, w}/{\rm erg\ s^{-1}} \in\{42, 51\}$. The distributions of $\beta_1$ and $\beta_2$ are assumed to be uniform in $\beta_1\in\{0.4, 1\}$ and $\beta_2\in \{1.0,\ 2.6\}$, and their values are randomly pick up in the ranges. Giving a parameter set \{$\beta_1$, $\beta_2$, $L_{b, w}$\}, we simulate a sample of $10^4$ dipole wind radiation events assuming that the dipole radiation wind is independent of the prompt emission jet. We measure the consistency of $L_{\rm w}$ and $z$ distributions between the mock and observed samples with the $K-S$ test. Figure \ref{fig:wind-LF-parameter} shows the $p_{\rm KS}$ contours in the $\beta_1-L_{b,w}$ and $\beta_2-\log L_{b, w}$ planes. In $1\sigma$ confidence level, we get $\beta_1=0.78^{+0.16}_{-0.15}$, $\log L_{b, w}/{\rm erg\ s^{-1}}=48.51^{+0.53}_{-0.65}$, and $\beta_2>1.6$. The best parameter set is $\{\beta_1, \beta_2, \log L_{b,w}\}=\{0.78, 2.22, 48.51\}$. Figure \ref{fig:wind-LF-Dis} shows the distribution probability contours of the mock sample derived from this parameter set in comparison with the observed data. We obtain $p_{\rm KS}=0.29$, indicating an good agreement between the two samples.

Observation shows the X-ray luminosity of the dipole radiation wind is tight correlated with prompt gamma-ray luminosity (e.g., \citealp{2019ApJ...877..153Z}). This motivates us to investigate the possible physical relation between them. We suspect that the wind is also structured and collimated, and make simulations based on the following assumptions. First, it is co-axial with the GRB jet and their viewing angle is same as the jet viewing angle. Second, its structure is also parameterized as Eq. \ref{jet_structure}, and parameters of the jet and wind are independent. Third, the luminosity ($L_{\rm c, w}$) within the core region ($\theta<\theta_{\rm c, w}$) is also assumed to be quasi-universal among bursts, which follows the Gaussian distribution with $\sigma_{\log L_{\rm c, w}}=0.5$. We constrain $L_{\rm c, w}$, $\theta_{\rm c, w}$ and the power-law index ($k_{\rm w}$) out of $\theta_{\rm c, w}$ using the same simulation technique as that for determining the jet structure parameters. The prior distributions of the parameters of the wind are taken as uniform distributions in the ranges of $k_{\rm w}\in \{2,10\}$, $\theta_{\rm c, w}\in\{0.5^{o},20^{o}\}$, and $\log L_{\rm c, w}/{\rm erg\ s^{-1}}\in \{42,\ 50\}$.

\begin{figure}

\plotone{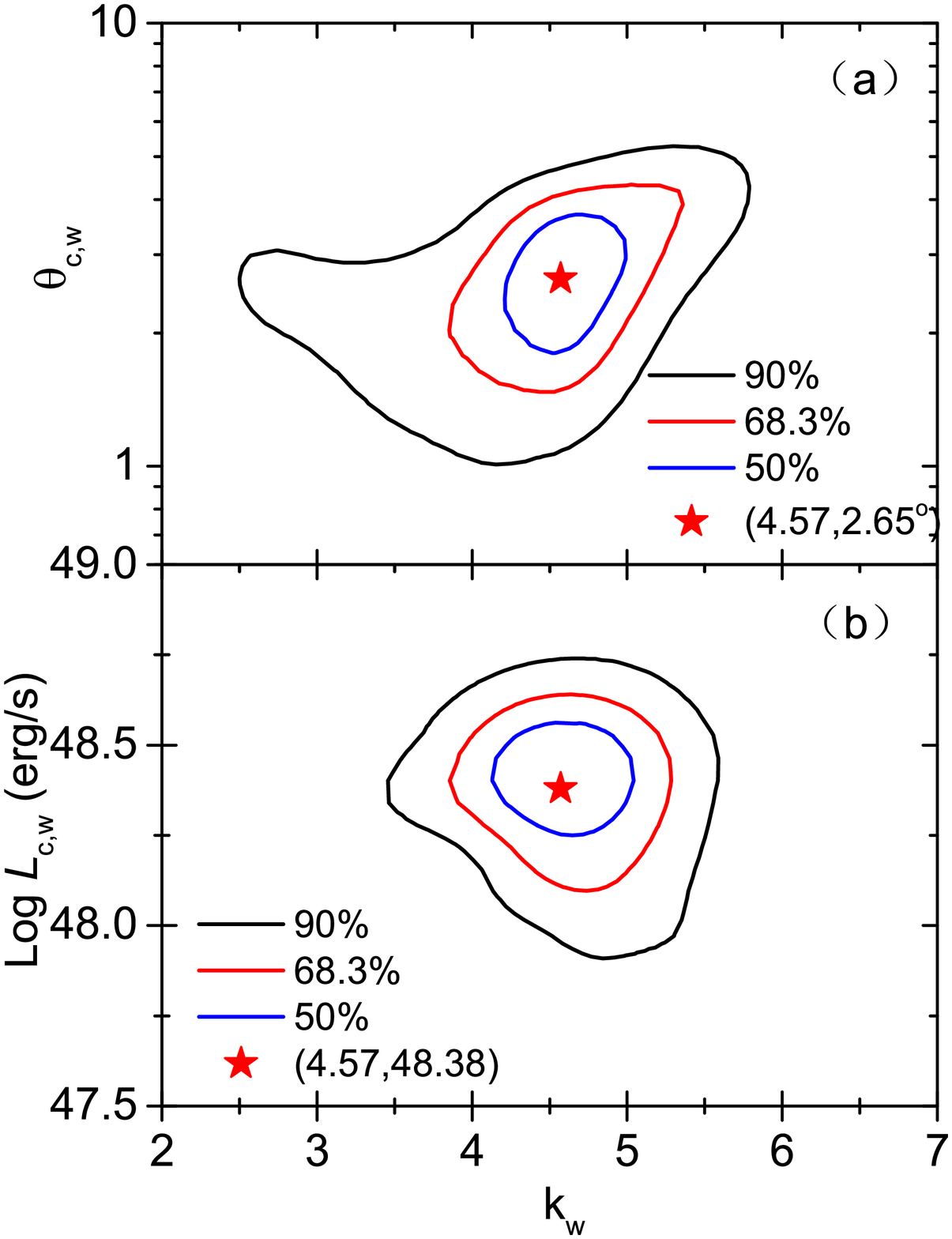}
\caption{Contours of the Kolmogorov-Smirnov test value $p_{\rm KS}$ ($p_{\rm KS}\equiv p^{L}_{\rm KS}\times p^{z}_{\rm KS}$) distribution in the $k_w-\log L_{\rm c, w}$ and $k_w-\theta_{c,w}$ planes, where $L_{\rm c, w}$ is the isotropic luminosity within $\theta_{c,w}$ and $k_w$ is the power-law decaying index out of $\theta_{c, w}$. The confidence level contours of 50\%, 68.3\% and 90\% are plotted by normalized the $P_{\rm KS}$ to their maximum value (the red stars).}
\label{fig:wind-structure}
\end{figure}

Figure \ref{fig:wind-structure} shows the $P_{\rm KS}$ contours in the $L_{\rm c, w}-k_{\rm w}$ and $\theta_{\rm c, w}-k_{\rm w}$ planes. The best parameter set is $\{\log L_{\rm c, w}, \theta_{\rm c, w}, k_{\rm w}\}=\{48.38, 2.65^{o}, 4.57\}$. They are well constrained even in a confidence level of $90\%$. In the $1\sigma$ confidence level, we obtain $\log L_{\rm c, w}/{\rm erg\ s^{-1}}=48.38^{+0.30}_{-0.48}$, $\theta_{\rm c, w}={2.65^{o}}_{-1.19^{o}}^{+0.1.73^{o}}$, and $k_{\rm w}=4.57^{+1.21}_{-0.75}$. It is found that $\theta_{\rm c, w}$ and $k_{\rm w}$ are comparable to $\theta_{\rm c, j}$ and $k_{\rm j}$, indicating that the wind structure is similar to the prior prompt emission jet.

\begin{figure}
\includegraphics[width=1\textwidth, angle=0]{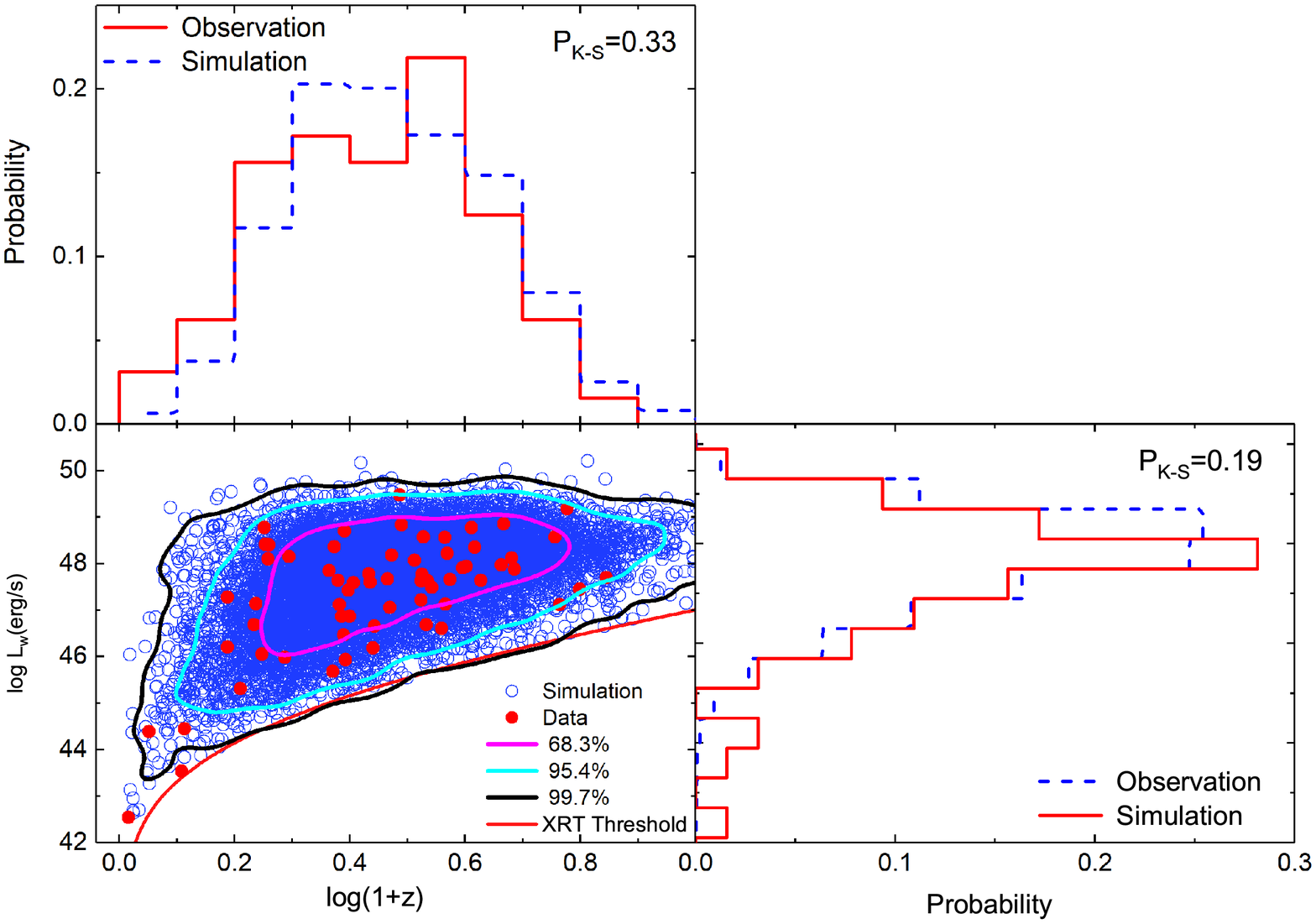}
\caption{Comparison between the observed sample (solid dots) and the probability distribution contours derived from our simulations based on the best parameter set $\{\log L_{\rm c, w}/{\rm erg\ s^{-1}}=48.38, \theta_{\rm c,w}=2.65^{o}, k_{\rm w}=4.57\}$. The solid red line stands for the sensitivity of $Swift$ XRT ($F_{\rm th}=1\times10^{-13}$ erg cm$^{-2}$ s$^{-1}$).}
\label{fig:wind-1D-2D}
\end{figure}

\begin{figure}
\includegraphics[width=1\textwidth, angle=0]{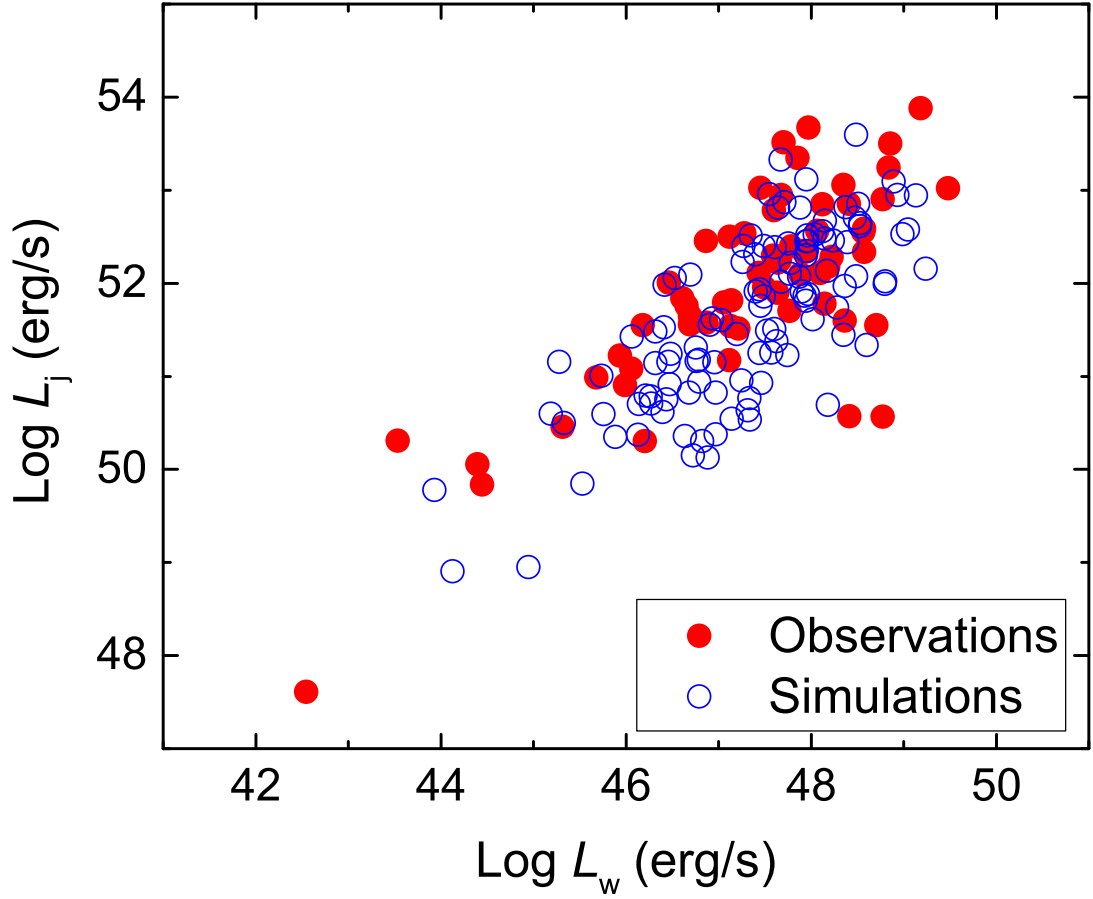}
\caption{Comparison between the observed sample (solid dots) and the simulations sample of $Swift$ based on the best parameter set of jet $\{\log L_{\rm c,j}/{\rm erg\ s^{-1}}=52.68\pm 0.05, \theta_{\rm c, j}=2.10^\circ, k_{\rm j}=4.00\}$ and wind  $\{\log L_{\rm c, w}/{\rm erg\ s^{-1}}=48.38, \theta_{\rm c,w}=2.65^{o}, k_{\rm w}=4.57\}$.
 }
\label{fig:Swift-Lw-Lj}
\end{figure}

We generate a mock sample of $10^4$ events with the best parameter set and shows the 1 and 2 dimensional (1-D and 2-D) distributions of $\log L_{\rm w}$ and $\log (1+z)$ in comparison with the data in Figure \ref{fig:wind-1D-2D}. By randomly picking up a sub-sample from this mock sample, we also show the $L_{\rm j}$ as a function of $L_{\rm w}$ in comparison with the data in Figure \ref{fig:Swift-Lw-Lj}. One can see that the observed 1-D and 2-D GRB distributions and the $L_{\rm j}-L_{\rm w}$ correlation can be reproduced in our simulations. The estimated detection probability of the low luminosity dipole radiation winds ($L_{\rm w}<10^{44}$ erg/s) in this wind structyre is $\sim 0.15\%$ with {\rm Swift}/XRT, and their average viewing angle and redshift are $\sim 24^{o}$ and 0.14, respectively.

\section{Discussion}

\subsection{Ejecta Structure and Physical Origin of Jet-Wind Connection}
 It was proposed that the geometrically-corrected jet energy among GRBs is quasi-universal in cases of a uniform jet or a quasi-Gaussian jet, i.e., $\sim 10^{51}$ ergs (\citealp{2001ApJ...562L..55F}; \citealp{2004ApJ...601L.119Z}). Our analysis suggests that the observed broad luminosity distribution of GRBs is resulted from observing quasi-universal jets at different viewing angles (see also \citealp{2016MNRAS.461.3607S}). The jet luminosity is $\sim 4.24\times 10^{49}$ erg s$^{-1}$,  and inferred jet energy is $4.24\times 10^{50}$ ergs based on the best parameter set from our analysis, assuming a burst duration of 10 seconds. This is comparable to that proposed by \cite{2001ApJ...562L..55F} and \cite{2004ApJ...601L.119Z}.

The derived jet structure in our analysis is composed of two components, a narrow constant luminosity core and a cocoon with power-lay-decay luminosity distribution. Signatures of two-component jet has been found in numerical simulations (e.g., \citealp{2003ApJ...586..356Z}). \cite{2004ApJ...601L.119Z} proposed that both GRB/XRF prompt emission/afterglow data can be described by a quasi--Gaussian-type structured jet with a typical opening angle of $\sim 6^{o}$ and a standard jet energy of $\sim 10^{51}$ ergs. \cite{2004MNRAS.348..153L} found that the prompt gamma-ray fluence as a function of jet opening angle can be derived as $S_{\gamma}\propto \theta^{-0.4}$ at $\theta<0.1$ rad and $S_{\gamma}\propto \theta^{-3.79}$ at $\theta>0.1$ rad. This is resemble to the structure obtained in our analysis. The observed broad distribution of the peak energy of the $\nu f_{\nu}$ spectrum may be also well explained with the two component jet model (\citealp{2004ApJ...608L...9L}; \citealp{2004ApJ...605..300H}). When the light of sight of an observer is within the narrow component, an observed burst is a typical GRB, and it may be an X-ray flash (XRF) if the light of sight is out of the core. Evidence of two-component jet is also seen in the afterglow lightcurves. For example, two different jet breaks are observed in early optical afterglow light curve (0.55 day, \citealp{2003Natur.423..844P}) and in late radio light curve (9.8 day) of GRB 030329. \cite{2003Natur.426..154B} proposed a two component jet model to explain the data. Millimeter observations of this burst further support the two--component jet model (\citealp{2003ApJ...595L..33S}).

Our analysis shows that the outflow of the wind may be also structured and collimated, and it is co-axial with the GRB jet. The derived $\theta_{\rm c, w}$ is slightly larger than $\theta_{\rm c, j}$, and $k_{\rm w}$ is also larger than $k_{\rm j}$. This result likely implies that the jet breaks out of the progenitor envelop, and the wind follows the pathway of the jet to form a collimated outflow. The interaction between the jet and medium result in a power-lay structured cocoon surrounding the narrow component. Since the jet and wind are in the same environment medium, the structures of the jet and wind ejecta would be analogue. Such a configuration incorporating the viewing angle effect may also explain the observed diversity of the X-ray and optical afterglow lightcurves.

\cite{2019ApJ...877..153Z} found that the energy releases of the jets and winds are correlated, i.e., $E_{\rm w}\propto E^{0.89\pm 0.07}_{\rm j}$ and proposed that the energy partition between the jet and wind is quasi-universal. Our simulations well reproduce the $L_{\rm w}-L_{\rm j}$ correlation , as shown in Figure \ref{fig:Swift-Lw-Lj}. This indicates that the observed jet-wind correlation would be attributed to the viewing angle effect to a quasi universal jet-wind configuration.

\subsection{Nature of LL-GRBs and Local GRB rate}
The local event rate ($\rho_0$) has been extensively discussed (e.g., \citealp{2001ApJ...552...36S}; \citealp{2005ApJ...619..412G} ;\citealp{2007ApJ...662.1111L}; \citealp{2010MNRAS.406.1944W}; \citealp{2014ApJ...783...24L}; \citealp{2019MNRAS.488.4607L}). The most uncertainty of $\rho_0$ is from whether the nearby LL-GRBs are the same population as typical GRBs. The inferred $\rho_0$ is usually about $\sim 1$ Gpc$^{-1}$ yr$^{-1}$ for typical high luminosity GRBs (\citealp{2001ApJ...552...36S}; \citealp{2007ApJ...662.1111L}; \citealp{2010MNRAS.406.1944W}). With observed nearby LL-GRBs 980425 and 060218, it was proposed that the local rate of LL-GRBs would be $\sim 10^3$ Gpc$^{-1}$ yr$^{-1}$, being much higher than typical GRBs (\citealp{2006ApJ...645L.113C}; \citealp{2007ApJ...662.1111L}; \citealp{2007ApJ...657L..73G}), and these LL-GRBs were suggested to be from a unique population (\citealp{2007ApJ...662.1111L}).

GRB 170817A is a {\em smoking gun} for extreme off-axis nature of LL-GRBs. To include GRB 170817A like extremely off-axis events ($\theta_v\sim 26^{o}$) in the mock sample, the best parameter set is $\{\log L_{\rm c, j}/{\rm erg\ s^{-1}}=52.68, \theta_{\rm c, j}=2.1^{o}, \ k_{\rm j}=4.00\}$. Note that LL-GRBs places strong constraints on the $k_{\rm j}$ value if both typical and low luminosity GRBs are the same population. The $k_j$ value is much shallower than that reported by  \cite{2015MNRAS.447.1911P}, who suggested $k_{\rm j}\sim -8$. The estimate of $\rho_0$ and local LL-GRB detection probability highly depends on the jet structure or the luminosity function of GRBs. Based on the derived jet structure parameters, we infer the detection probability of GRBs 060218 and 170817A like extreme off-axis LL-GRBs with {\em Swift}/BAT is about $10^{-4}$, and $\rho_0\sim 9.6$ Gpc$^{-1}$ yr$^{-1}$. The $\rho_0$ value is about one order magnitude larger than that derived from typical GRB samples and about two orders of magnitude smaller than that estimated with a few nearby LL-GRBs.


\subsection{GRB and Magnetar Dipole Radiation Survey with the {\em Einstein Probe} (EP)}
The Einstein Probe (EP) is dedicated to make X-ray transient survey and monitor variable objects in the soft X-ray band (0.5-4keV). A wide field X-ray telescope (WXT) and a follow-up X-ray telescope (FXT) are on board EP. The field-of-view of WXT is about 1.1 steradian and its spatial resolution is about 5 arcmin. Using the novel lobster-eye optics, WXT may offer high sensitivity for X-ray all-sky monitors. Its sensitivity is bout $10^{-11}$ ergs s$^{-1}$ cm$^{-2}$ in the 0.5-4 keV band (or ~0.3 mCrab) at the $5\sigma$ confidence level in a 1000s exposure (\citealp{2016SSRv..202..235Y}). The appropriate exposure time would hundreds of seconds since the typical spin down timescale of newly-born magnetars is about several thousands seconds. In a 100s exposure, WXT sensitivity is about $10^{-10}$ ergs s$^{-1}$ cm$^{-2}$, which is higher by one or two orders of magnitude than Swift/BAT. FXT aims at performing follow-up characterization and precise localization of newly-discovered transients. Its sensitivity is $F^{\rm FXT}_{\rm th}=3 \times 10^{-11}$ ergs s$^{-1}$ cm$^{-2}$ in 100s exposure.

\begin{figure}
\includegraphics[width=1\textwidth, angle=0]{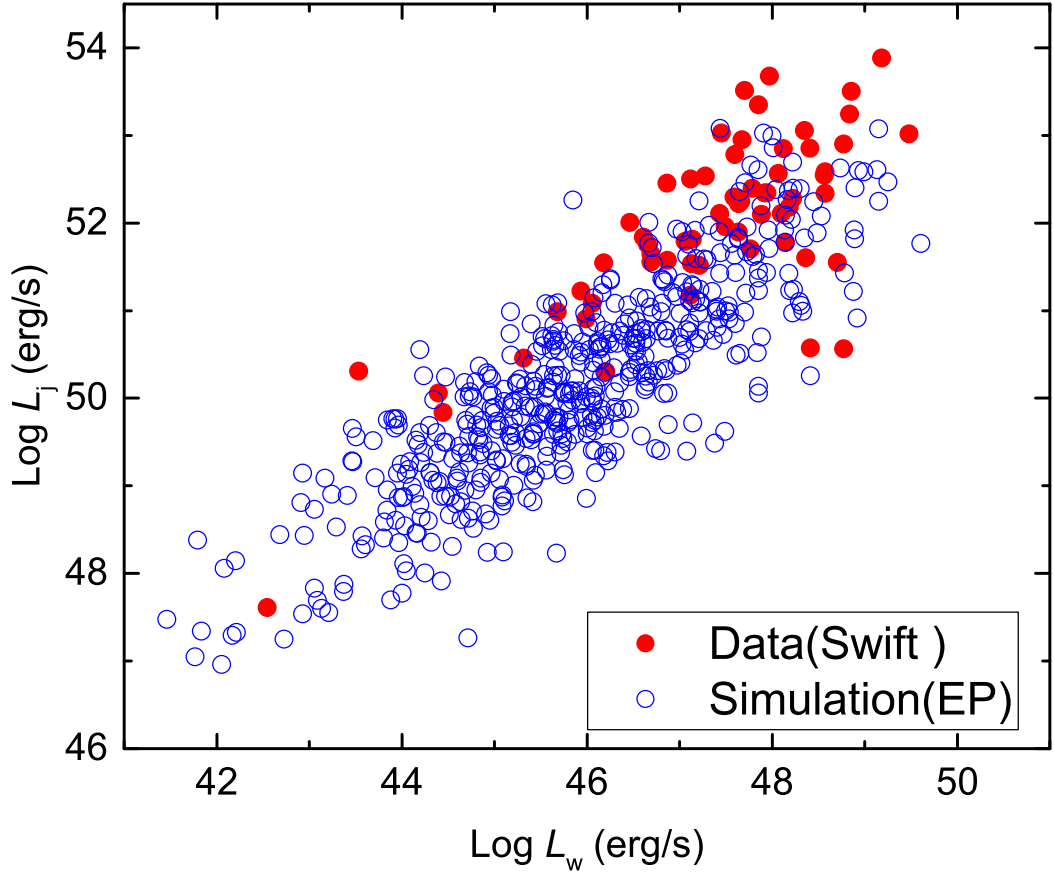}
\caption{Simulated $L_{\rm j}-L_{\rm w}$ correlation for the GRBs detected with the {\em EP} mission (open circles) in one operation year in comparison with the current sample observed with the {\em Swift} mission (red dots). Our simulations are based on the best structure parameters of the jet and wind and the local GRB rate derived in this analysis. The X-ray emission in the {\em EP}/WXT band from the jet and wind components are taken into account in evaluation of the WXT trigger. Redshift measurement and the effects of the wind covering by jet afterglows as well as other observation al biases are not considered.
\label{fig:EP-Lw-Lj}}
\end{figure}

\begin{figure}
\includegraphics[width=1.0\textwidth, angle=0]{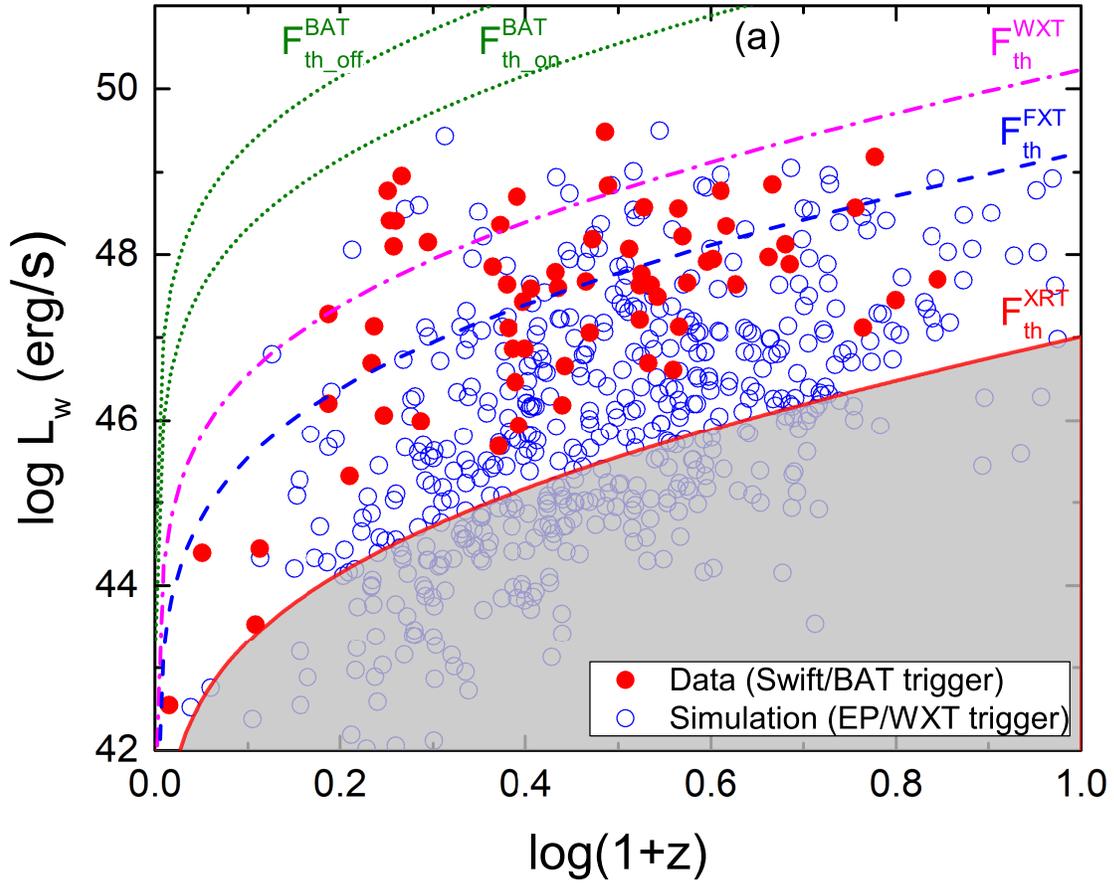}
\caption{Simulations of GRBs triggered by {\em EP}/FXT in one operation year in the $\log (1+z)-\log L_{\rm w}$ plane based on the best structure parameters of the jet and wind and the local GRB rate derived in this analysis. Thresholds of different instruments are also marked for examining the detection probability of their wind emission with these instruments.
\label{fig:EP-Lw-z}}
\end{figure}

We investigate the possible survey outcomes for the dipole radiations with the WXT and FXT on board the EP mission. We simulate the jet and wind emission of a GRB and evaluate whether it can be triggered with WXT and its wind emission can be detectable with FXT based on the derived best parameter sets of the jet and wind structure as well as the local GRB rate. Adopting the sensitivity of EP/WXT as $F^{\rm WXT}_{\rm th}=1\times 10^{-10}$ ergs s$^{-1}$ cm$^{-2}$ in 100s exposure, we find that WXT may trigger 510 events per year, which is higher than {\em Swift}/BAT with a factor of about 5. This sample is rich of XRFs or LL-GRBs since WFX is sensitive in soft X-ray band. The $L_{\rm j}-L_{\rm w}$ correlation thus may be further verified with this sample, as shown in Figure \ref{fig:EP-Lw-Lj}. The great challenge is detection of the wind emission. Figure \ref{fig:EP-Lw-z} shows this sample in the $\log (1+z)-\log L_{\rm w}$ plane. One can find that the wind radiations is detectable with WXT is only 2.4\% (about 12 GRBs per year). Prompt follow-up observation with FXT may increase the percentage to $8.6\%$ (about 44 GRBs per year). A large fraction of the wind events should be detectable with XRT. However, as mentioned in \cite{2019ApJ...877..153Z}, the major factor hampering the identification of the wind radiations is the afterglows, not the XRT sensitivity. They found that about $60\%$ MD radiations may be covered by the forward shock afterglows. Although the WXT and FXT detectable wind events are bright, the covering effect by bright X-ray afterglows would also significantly reduce the detection rate of the wind events.

\cite{2019Natur.568..198X} reported an orphan magnetar-powered X-ray transient using the data observed with Chandra X-ray telescope. Within the jet-wind configuration in this analysis, the non detection of jet radiation accompanied the orphan X-ray transient is not due to the viewing angle effect since the jet and wind are co-axial. It is possible that the GRB jet as a pioneer does not successfully break out and the wind as a successor breaks out as a collimated outflow to produce the X-ray transient.
Thus, WXT survey may be also interesting for searching similar orphan X-ray transients. Since the wind emission is steady and long last, the trigger algorithm is critical to trigger such kind of events.

\section{Conclusion}
By assuming that GRB jets powered by newly-born magnetars are quasi-universal and the observed broad luminosity distribution is due to the viewing angle effect, we have constrained the GRB jet structure through Monte Carlo simulations.  We parameterize the jet structure $L_{\rm j}=L_{\rm c,j}$ at $\theta\leqslant \theta_{\rm c, j}$ and $L_{\rm j}=L_{\rm c, j}(\theta/\theta_c)^{-k_{\rm j}}$ at $\theta>\theta_{\rm c, j}$, and constrain the parameters by comparing our simulations to the {\em Swift} data via MCMC simulations. We find $\log L_{\rm c, j}/{\rm erg\ s^{-1}}=52.68^{+0.76}_{-0.33}$ and $k_{\rm j}=4.00^{+0.27}_{-0.37}$ in $1\sigma$ confidence level, $\theta_{\rm c, j}=2.10_{-1.28}^{+1.90}$ in a confidence level of $50\%$, and the inferred local GRB rate as $\rho=9.6$ Gpc$^{-3}$ yr$^{-1}$ by including both the typical GRBs and LL-GRBs as the same population. The typical $\theta_v$ is 3.3$^{o}$ for typical GRBs, and they may be $20^{o}\sim30^{o}$ for LL-GRBs, which are detectable at nearby universe with a chance probability of $\sim 10^{-4}$. The X-ray luminosity function of the dipole radiation wind can be empirically described with a broken power-law function with indices $\beta_1=0.79$ and $\beta_2<1.6$ broken at $3.2\times 10^{48}$ erg s$^{-1}$. We further investigate the scenario that the wind outflow may be also collimated and co-axial with the prompt GRB jet. It is found that structure of wind is similar to the prior prompt emission jet, i.e., $\log L_{\rm c, w}/{\rm erg\ s^{-1}}=48.38^{+0.30}_{-0.48}$, $\theta_{\rm c, w}={2.65^{o}}_{-1.19^{o}}^{+0.1.73^{o}}$, and $k_{\rm c, w}=4.57^{+1.21}_{-0.75}$ in the $1\sigma$ confidence level. The observed correlation between the prompt gamma-ray luminosity of the jet and X-ray luminosity of the wind, indicating that the correlation may be resulted from the viewing effect to a quasi universal structured ejecta of both the jet and wind component.

Our analysis is based on a sub-sample of GRBs whose central engine may be a newly-born magnetar. Their early X-ray lightcurves are characterized with a plateau or a shallow decay segment. Sample selection with such a feature may suffer great biases since the early bright X-ray afterglows of the GRB jet, erratic late X-ray flares from late activities of the GRB central engine, the X-ray tail emission of prompt gamma-rays from high latitude of the GRB fireball may make a lot contaminations to this phase (e.g., \citealp{2019ApJ...877..153Z}). As mentioned in \cite{2019ApJ...877..153Z}, the major factor hampering the identification of the wind radiations is the jet afterglows. They found that the plateau may be covered by jet X-ray afterglows form the forward shocks for about $60\%$ GRBs. We select only those GRBs that their plateau or the shallow decay segment is clear detected without significant contaminations as mention above. The luminosity and redshift distributions of this sub-sample do not show significant different from the entire sample of the {\em Swift}/BAT GRBs. We regard it as a uniform sub-sample picked up from the complete sample under the {\em Swift}/BAT threshold in this analysis. Larger uniform sample, especially those samples including more low or median luminosity GRBs with $L_{j}\sim 10^{46}-10^{49}$ erg s$^{-1}$, may place stronger constraints on the jet and wind structure parameters as well as the local event rate of GRBs. However, we should emphasize that the fundamental issue of the sample selection effect that may dramatically change our analysis results is whether LL-GRBs are the same population as typical GRBs. 

Redshift measurement is necessary for our analysis. It suffers great observational biases, including the optical afterglow detection, optical spectroscopic observations of the afterglow or host galaxy identification (e.g., \citealp{2010MNRAS.406..558Q}). In addition, the derived jet structure in our analysis is resemble to a narrow uniform jet surrounding by a cocoon featured as $L{_{\rm j}}\propto (\theta/\theta_{\rm c, j})^{-4}$ at $\theta>\theta_{\rm c, j}$. The cocoon luminosity rapidly dimer as $\theta_v$ increase. Thus, the observed afterglows from an observer with a light of sight out of the $\theta_c$ may be dim and onset later (e.g., \citealp{2008MNRAS.387..497P}; \citealp {2009A&A...499..439G}; \citealp{2010MNRAS.402...46M}). This should also make a bias for measurement of their redshifts.

\acknowledgments
We acknowledge the use of the public data from the {\em Swift} data archive and the UK {\em Swift} Science Data Center. This work is supported by the National Natural Science Foundation of China (Grant No.11533003, 11851304, 11603006, and U1731239), Guangxi Science Foundation (grant No. 2017GXNSFFA198008, 2016GXNSFCB380005 and AD17129006), the One-Hundred-Talents Program of Guangxi colleges, the high level innovation team and outstanding scholar program in Guangxi colleges, Scientific Research Foundation of Guangxi University
(grant No. XGZ150299), and special funding for Guangxi distinguished professors (2017AD22006).

\clearpage

\begin{deluxetable}{lcccll}
\tabletypesize{\scriptsize}
\tablewidth{0pt}
\tablecaption{The GRB data sample.\label{tab:TDE}}
\tablehead{$GRB$  & $z$ & $\Gamma_{\rm BAT}$ & $P^{a}$ (ph.cm$^{-2}$.s$^{-1}$) & $\log L_{\rm w}$ (erg/s)  & $\log L_{\rm j}$ (erg/s)}
\startdata
050315 & 1.949 & 2.16 & 1.93 & 47.06 & 51.79 \\
050319 & 3.24 & 2.11 & 1.52 & 47.64 & 52.22 \\
050802 & 1.71 & 1.30 & 2.62 & 47.78 & 52.40 \\
050814 & 5.3 & 1.21 & 0.71 & 47.45 & 53.03 \\
050822 & 1.434 & 2.11 & 2.43 & 46.87 & 51.58 \\
050826 & 0.297 & 1.24 & 0.39 & 44.45 & 49.84 \\
051016B & 0.9364 & 1.94 & 1.30 & 45.99 & 50.91 \\
060202 & 0.785 & 1.59 & 0.51 & 48.77 & 50.57 \\
060204B & 2.3393 & 1.43 & 1.35 & 47.63 & 52.29 \\
060502A & 1.51 & 0.97 & 1.69 & 46.86 & 52.45 \\
060604 & 2.68 & 1.75 & 0.34 & 47.13 & 51.54 \\
060605 & 3.8 & 0.90 & 0.47 & 48.12 & 52.85 \\
060607A & 3.082 & 1.13 & 1.40 & 48.77 & 52.90 \\
060614 & 0.125 & 1.62 & 11.51 & 44.39 & 50.06 \\
060714 & 2.71 & 1.61 & 1.41 & 48.22 & 52.28 \\
060729 & 0.54 & 2.10 & 1.41 & 46.20 & 50.31 \\
060814 & 1.9229 & 1.31 & 7.27 & 47.67 & 52.95 \\
061121 & 1.314 & 1.03 & 21.08 & 47.85 & 53.35 \\
061202 & 2.253 & 1.40 & 2.55 & 48.07 & 52.56 \\
070110 & 2.352 & 1.68 & 0.60 & 47.77 & 51.70 \\
070129 & 2.3384 & 1.97 & 0.59 & 47.21 & 51.52 \\
070306 & 1.496 & 1.65 & 4.25 & 47.43 & 52.11 \\
070328 & 2.0627 & 1.10 & 4.22 & 49.48 & 53.02 \\
070508 & 0.82 & 1.09 & 24.68 & 48.41 & 52.86 \\
070521 & 2.0865 & 1.09 & 6.71 & 48.83 & 53.24 \\
080310 & 2.43 & 1.97 & 1.30 & 47.63 & 51.90 \\
080430 & 0.767 & 1.80 & 2.65 & 46.06 & 51.08 \\
080905B & 2.374 & 1.48 & 1.66 & 48.57 & 52.34 \\
081029 & 3.847 & 1.66 & 0.46 & 47.88 & 52.10 \\
090404 & 3.0 & 1.85 & 1.94 & 47.94 & 52.34 \\
090407 & 1.4485 & 0.96 & 0.64 & 46.46 & 52.01 \\
090529 & 2.625 & 1.70 & 0.65 & 46.61 & 51.84 \\
090618 & 0.54 & 1.15 & 38.04 & 47.28 & 52.54 \\
091018 & 0.971 & 2.19 & 10.28 & 48.15 & 51.78 \\
091029 & 2.752 & 1.80 & 1.78 & 47.66 & 52.25 \\
100302A & 4.813 & 1.47 & 0.48 & 47.12 & 52.50 \\
100418A & 0.6235 & 2.63 & 1.04 & 45.31 & 50.46 \\
100425A & 1.755 & 2.32 & 1.36 & 46.18 & 51.55 \\
100615A & 1.398 & 1.56 & 5.43 & 47.64 & 52.23 \\
100704A & 3.6 & 0.99 & 4.30 & 47.97 & 53.68 \\
100901A & 1.408 & 1.84 & 0.79 & 47.11 & 51.17 \\
100906A & 1.727 & 1.50 & 10.14 & 47.60 & 52.79 \\
110213A & 1.46 & 2.63 & 1.56 & 48.70 & 51.55 \\
110808A & 1.348 & 1.61 & 0.38 & 45.68 & 50.99 \\
111008A & 4.9898 & 1.26 & 6.42 & 49.18 & 53.88 \\
111228A & 0.714 & 1.85 & 12.43 & 46.69 & 51.65 \\
120118B & 2.943 & 1.90 & 2.16 & 47.91 & 52.35 \\
120422A & 0.283 & 0.94 & 0.55 & 43.53 & 50.31 \\
120521C & 6.0 & 1.27 & 1.90 & 47.70 & 53.51 \\
120811C & 2.671 & 1.85 & 4.08 & 48.56 & 52.55 \\
121027A & 1.773 & 1.67 & 1.29 & 46.65 & 51.76 \\
140512A & 0.725 & 1.39 & 6.80 & 47.14 & 51.82 \\
140518A & 4.707 & 1.74 & 1.01 & 48.57 & 52.58 \\
140703A & 3.14 & 1.29 & 2.85 & 48.35 & 53.06 \\
141121A & 1.47 & 1.93 & 0.89 & 45.93 & 51.22 \\
150910A & 1.359 & 1.45 & 1.06 & 48.36 & 51.60 \\
151027A & 0.81 & 1.24 & 6.86 & 48.10 & 52.11 \\
161117A & 1.549 & 1.72 & 6.83 & 47.59 & 52.30 \\
170113A & 1.968 & 1.28 & 1.08 & 48.18 & 52.18 \\
170202A & 3.645 & 1.21 & 4.74 & 48.85 & 53.50 \\
170714A & 0.793 & 1.49 & 0.40 & 48.41 & 50.57 \\
171205A & 0.0368 & 1.98 & 0.95 & 42.54 & 47.61 \\
171222A & 2.409 & 2.13 & 0.67 & 46.69 & 51.56 \\
180115A & 2.487 & 1.45 & 0.58 & 47.49 & 51.96 \\
\enddata
\tablenotetext{a}{ The 1 s peak photon flux.}
\end{deluxetable}

\clearpage


\clearpage

\clearpage

\clearpage

\clearpage
\clearpage

\clearpage

\clearpage

\clearpage

\clearpage

\end{document}